\documentclass[10pt,journal,doublecolumn,singlespacing]{IEEEtran}
\usepackage{url}
\usepackage{tocloft}
\usepackage{setspace}
\usepackage{amsmath}
\usepackage{multicol}
\usepackage{amssymb}
\usepackage[utf8]{inputenc}
\usepackage[T1]{fontenc}
\usepackage[font={small}]{caption}
\usepackage{color, soul}
\usepackage{cite}
\usepackage{enumerate}
\usepackage[square, comma, numbers, sort&compress]{natbib}
\usepackage{subfig}
\usepackage{bm}
\usepackage{multirow}
\usepackage{tabularx}
\usepackage{subfig}
\usepackage{upgreek}
\usepackage{makecell}

\usepackage{url}
\usepackage{tocloft}
\usepackage{setspace}
\usepackage{amsmath}
\usepackage{multicol}
\usepackage{amssymb}
\usepackage[utf8]{inputenc}
\usepackage[T1]{fontenc}
\usepackage[font={small}]{caption}
\usepackage{cite}
\usepackage{enumerate}
\usepackage[square, comma, numbers, sort&compress]{natbib}
\usepackage{subfig}
\usepackage{bm}
\usepackage{multirow}
\usepackage{algorithm}
\usepackage{algorithmic}
\usepackage{float}

\makeatother

\providecommand{\theoremname}{Theorem}

\makeatother

\providecommand{\lemmaname}{Lemma}

\makeatother

\providecommand{\propositionname}{Proposition}

\makeatother

\providecommand{\corname}{Corollary}

\makeatother

\providecommand{\remname}{Remark}

\ifCLASSINFOpdf
\else
\fi
\ifCLASSINFOpdf
\usepackage[pdftex]{graphicx}
\usepackage{epstopdf}
\else
\fi

\begin{document}
	
	\title{Perpetual Reconfigurable Intelligent Surfaces Through In-Band Energy
		Harvesting: Architectures, Protocols, and
		Challenges}

	\author{Konstantinos Ntontin, \IEEEmembership{Member, IEEE},
		{Alexandros--Apostolos A. Boulogeorgos}, \IEEEmembership{Senior Member, IEEE}, 
		Sergi Abadal, \IEEEmembership{Member, IEEE}, Agapi Mesodiakaki
		\IEEEmembership{Member, IEEE}, Symeon Chatzinotas, \IEEEmembership{Fellow,
			IEEE}, and Bj\"{o}rn Ottersten, \IEEEmembership{Fellow, IEEE}}
	
	
	
	
	
	

\maketitle

\begin{abstract}
	Reconfigurable intelligent surfaces (RISs) are considered to be a key
	enabler of highly energy-efficient 6G and beyond networks. This property arises
	from the absence of power amplifiers in the structure, in contrast to active
	nodes, such as small cells and relays. However, still an amount of power is required for their operation. To improve their energy efficiency further, we propose the notion of \emph{perpetual} RISs, which secure the power needed to supply their
	functionalities through wireless energy harvesting of the impinging transmitted
	electromagnetic signals. Towards this, we initially explain the rationale behind
	such RIS capability and proceed with the presentation of the main RIS controller architecture
	that can realize this vision under an in-band energy harvesting consideration.
	Furthermore, we present a typical energy-harvesting architecture followed by two
	harvesting protocols. Subsequently, we study the performance of the two protocols under a typical communications scenario. Finally, we elaborate on the main research challenges governing the realization of large-scale networks with perpetual RISs.
\end{abstract}

\section{Introduction}
\label{Introduction}

Although using millimeter-wave (mmWave) bands, in order to prevent
the capacity crunch of sub-6 GHz bands, has been envisioned and standardized for
5G networks, wide-scale network deployment on these bands is expected to be realized
in their 6G counterparts. The large bandwidth offered in mmWave bands is
essential not only for boosting the communication rates, but also for achieving
sub-meter localization that is required in several challenging use cases with a high societal impact, such
as autonomous driving in urban areas
\cite{Requirements_autonomous_vehicle_navigation}, highly accurate localization
of Intenet of Things (IoT) devices in a smart factory
\cite{Overview_Integrated_Localization_Paper}, and indoor navigation of people
with impaired vision \cite{Indoor_Navigation_Visual_Impairments}. 

However, mmWave bands are more susceptible to fixed and moving blockages in comparison with their sub-6-GHz counterparts. A straightforward solution to counteract this
bottleneck is the large network densification with small cells and relays so
that line-of-sight (LoS) connections between them and the end users are achieved
with very high probability. However, such a solution may be prohibitive from a
cost and energy consumption point of view
\cite{Khawaja_mmWave_Passive_Reflectors}.

To counteract the aforementioned bottleneck, reconfigurable intelligent surfaces
(RISs) are widely believed to be a viable alternative due to their
capability for conformal designs and notably lower power consumption compared
with active nodes that are equipped with power amplifiers (PAs). This is due to
the lack of PAs in the RIS case, which is the most power consuming electronic
component \cite{Cunhua_RIS_6G}. By adjusting the impedance of their unit cells
(UCs), RISs are able to perform a variety of functions, such as reflection,
absorption, diffraction, and polarization change of the incident electromagnetic
wave. Owing to their ease of deployment, RISs are expected to be ubiquitously
deployed in both indoor and outdoor scenarios in the forthcoming 6G and beyond
networks, especially for mmWave bands, so as to provide numerous alternative
transmitter-RIS and RIS-receiver LoS routes in case of blockages. They can
assist not only communications, but also localization simultaneously
\cite{Jiguang_He_RIS_Simultaneous_Communications_Localization}. 

\subsection*{Why Do We Need Perpetual RISs?}

Current RIS prototypes base their
reconfigurability on field-programmable gate array (FPGA) controllers that
normally exhibit power consumption levels that require the RIS to be constantly
plugged onto the power grid. This need could impair the requirement for
a pervasive RIS deployment due to difficulty of massively wiring them to the grid. In particular, deploying cables
involves planning and notable maintenance costs that can immensely grow for
massive deployments \cite{Flexibility_cost_energy_efficiency_smart_home}.
Moreover, requests to local authorities for the permission of installing the
required wired infrastructure would be needed in several occasions, which are
usually time consuming processes. In addition, there are places that the power
grid would not be allowed to reach to for preventing urban visual pollution.

Additionally, supplying the energy needs of RISs with single-use batteries is
also not a viable solution either because this would give rise to a large effort
for regular replacements of a massive number of single-use batteries, let alone
the constant monitoring of their level that would be required. Based on the aforementioned powering issues that
a massive RIS deployment would induce, the following question arises:
\emph{\textbf{Could RISs perpetually operate by means of wireless energy
		harvesting from the impinging electromagnetic (EM) signals that are used for
		communications and localization.}}

In the remainder of this article, we first present the two main RIS controller
architectures and explain why only one of these can potentially result in
perpetual operation. Subsequently, we introduce an energy-harvesting (EH) architecture together with
two in-band EH protocols. Furthermore, their performance is compared. Finally,
we identify a number of research challenges for the realization of perpetual
RISs and conclude this article with the main takeaways.

\section*{Controller Architectures}

\begin{figure}
	\label{RIS_controllers}
	\centering
	{\includegraphics[width=3in, height=3in]{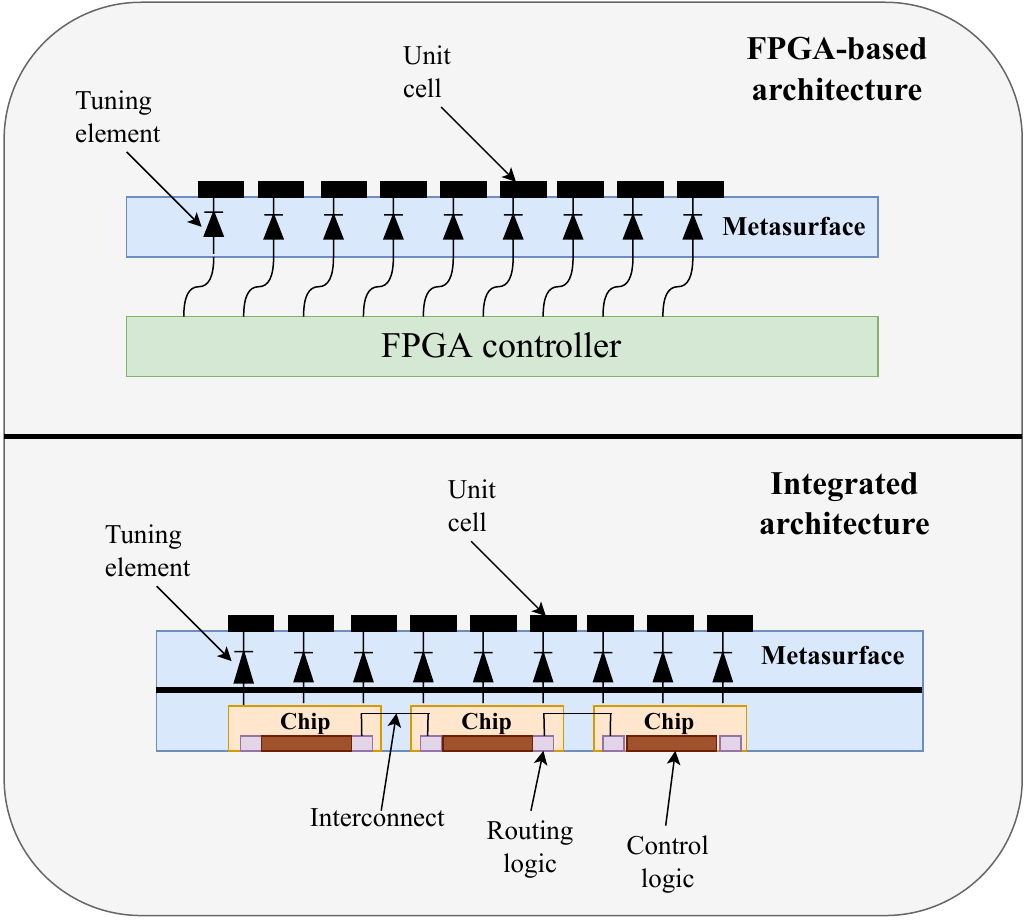}}
	\caption{Controller architectures for RISs
		\cite{Abadal_2020_programmable_metamaterials}.}
	\label{RIS_controllers}	
\end{figure}

Let us now present the two basic RIS controller  architectures, namely the
conventional \emph{FPGA-based architecture} and the \emph{integrated
	architecture}. In addition, we will elaborate on why the integrated architecture
is the only viable approach for perpetual RIS operation.

\subsection*{FPGA-based architecture}

As depicted in Fig.~\ref{RIS_controllers}, in this architecture the FPGA acts as
an external controller and adjusts the bias voltages of the tuning elements that
are attached to the UCs. This, in turn, alters the impedance of the UCs so that
the desired metasurface function is realized. The tuning elements normally
comprise varactors or variable resistors, positive intrinsic negative diodes or
switches, microelectromechanical systems, mechanical parts, or advanced
materials, such as graphene, or liquid crystals
\cite{Abadal_2020_programmable_metamaterials}.
The FPGA-based architecture is the conventional architecture with which several
proof-of-concept RISs have been designed and manufactured. It offers the
advantage of separate design of the metasurfaces and FPGAs. On the other hand,
FPGA-based architectures are usually bulky and exhibit a significant power
consumption that make perpetual operation challenging
\cite{Abadal_2020_programmable_metamaterials}.  

\subsection*{Integrated architecture}

In contrast to the FPGA-based architecture, the integrated architecture relies
on the integration of a network of communicating chips within the metasurface
containing tuning elements, control circuits, and even sensors. As it is pointed
out in \cite{Abadal_2020_programmable_metamaterials}, integrated architectures
are custom-made and are therefore much more optimized than FPGA-based
architectures. This means that the control sub-system is less intrusive in terms
of EM interference, less bulky, and potentially exhibits lower power
consumption. Hence, perpetual operation is envisioned as a possibility for the
integrated RIS architecture by means of wireless energy harvesting
\cite{Abadal_2020_programmable_metamaterials}. The metasurface controlling
chips, that would wirelessly receive reconfiguration commands under perpetual
operation,  may consist of circuitry that reads the UC state and
digital-to-analog (DAC) converters that adjust the bias voltage to the tuning
elements. 

\begin{figure}
	\label{ASIC_diagram}
	\centering
	{\includegraphics[width=2in, height=2in]{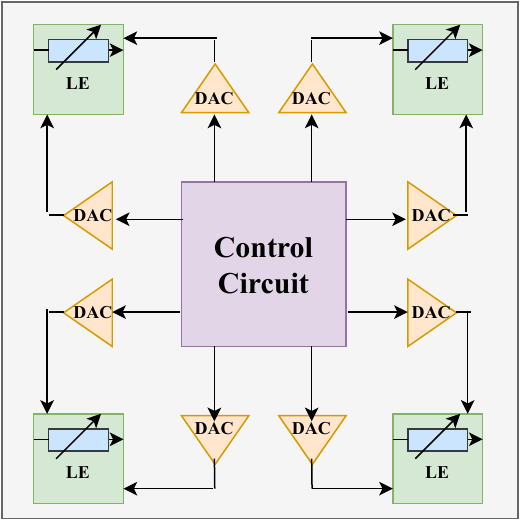}}
	\caption{Top-level diagram of the ASIC used in \cite{Julius_ASIC_Low_Power} as
		the controling chip.}
	\label{ASIC_diagram}	
\end{figure}

A possible architecture for such a controlling chip, based on
application-specific integrated circuits (ASICs) for simultaneously controlling
the response of $4$ UCs, is depicted in Fig.~\ref{ASIC_diagram}
\cite{Julius_ASIC_Low_Power}. According to the particular example\footnote{The actual ASIC design may change based on the application and the type of UCs used.}, the ASIC comprises: i) the control circuit, ii)
the DACs, and iii) the radio frequency (RF) tunable loading elements (LEs). The control
circuit is responsible for the communication operations of the ASIC by
wirelessly receiving reconfiguration commands\footnote{We assume that a wireless
	receiver is embedded into the control circuit.} and sending/receiving
communication data to/from its neighboring controllers. In the particular
implementation of \cite{Julius_ASIC_Low_Power}, the control circuit consists of
an internal memory with 64 cells that store the reconfiguration data that are
required by the LEs for adjusting the impedance of the UCs. In addition, the
control circuit integrates another internal memory with 18 cells for storing the
data for networking among the ASICs. In turn, the cells that store the RIS
reconfiguration data drive the inputs of the $8$ DACs. Furthermore, the output
of the DACs drives the input of the LEs. The LEs consist of a 
metal–oxide–semiconductor field-effect transistor (MOSFET) varistor that adjusts
the real part of the UC impedance and a MOSFET varactor that adjusts its
imaginary part. Finally, we note that an important feature of the ASIC proposed
in \cite{Julius_ASIC_Low_Power} is its asynchronous  operation, which can result
in a notably lower circuit consumption compared with a synchronous
implementation.

\section{Energy Harvesting Architecture, Power Consumption Model, and Proposed
	Harvesting protocols} \label{Section_II}

In this section, we first present a typical EH architecture for supplying the
energy needs of a perpetual RIS. Next, we introduce the power consumption model
based on the considered integrated architecture for reconfiguring the surface.
Finally, we report two harvesting protocols based on either time- or
UC-splitting. 

\subsection*{Energy Harvesting Architecture}

\begin{figure}
	\label{Harvesting_Architecture}
	\centering
	{\includegraphics[width=3.4in,
		height=1.4in]{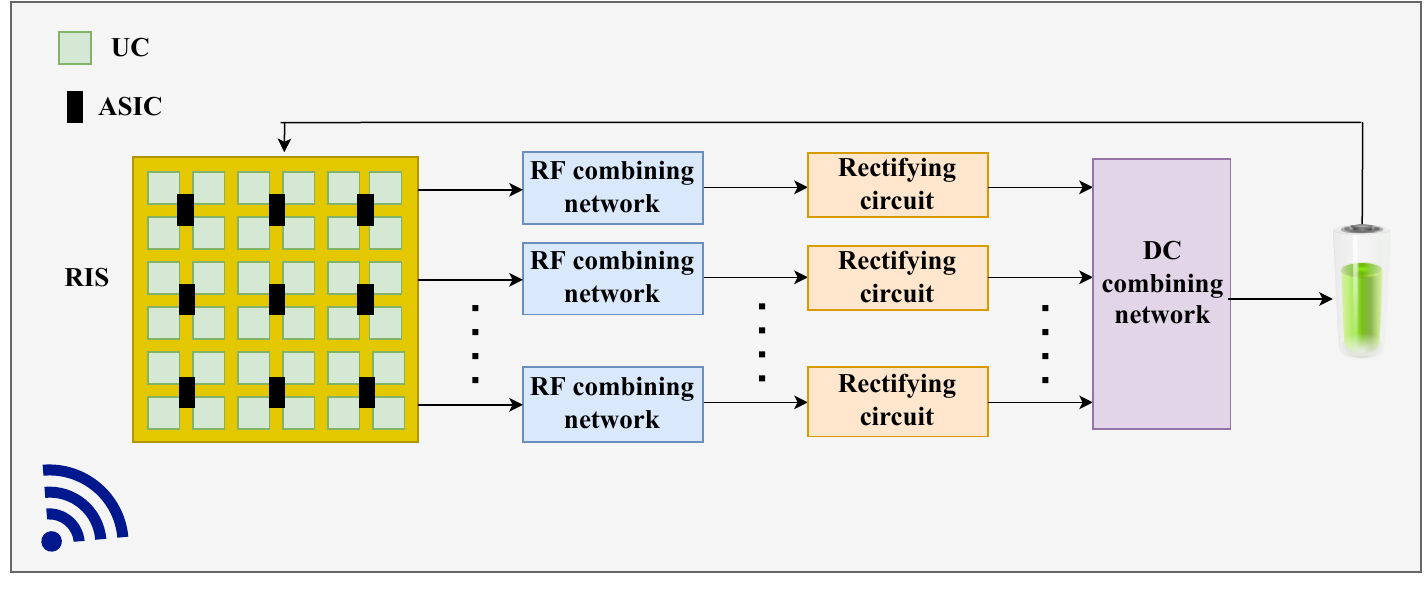}}
	\caption{Energy harvesting architecture.}
	\label{Harvesting_Architecture}	
\end{figure}

The EH architecture is depicted in Fig.~\ref{Harvesting_Architecture}. The
absorbed power of subsets of UCs is combined in the RF domain and the combined
outputs drive an equal number of rectifying circuits that transform the RF
energy to a direct current (DC) one. A DC combining network combines the DC
powers and its output charges a battery that is used to power the ASICs.

The presented architecture is a compromise between the two extreme cases of: i)
combining in the RF domain the absorbed powers of all the UCs and ii) enabling
each UC to drive a single rectifying circuit. The first case may result in
substantial insertion losses due to RF combining, if the absorbed powers are not
perfectly phase aligned, whereas the absorbed power of each UC in the second
case might not be sufficient to turn on the rectifying circuit
\cite{Energy_harvesting_metasurface_comprehensive_review}. Hence, the 
architecture presents a flexible design and the amount of chains is subject to
optimization, based on the specific application and electronic packaging
considerations. Finally, as far as the rectifying circuit is concerned, which is
a passive device, the three main options for its realization are a diode, where
a Schottky diode is the most common implementation, a bridge of diodes, and a
voltage rectifier-multiplier
\cite{Energy_harvesting_metasurface_comprehensive_review}.

\subsection*{Power Consumption Model}

Due to the fact that the RF/DC combiners and rectifying circuits in the
presented energy harvesting architecture are passive devices, the only source of
power consumption in the RIS is the ASIC. As with any electronic device, this
power consumption consists of the summation of a static and a dynamic part. The
latter part is due to the wireless reception of reconfiguration commands, the
switching operations and the resulting charging/discharging of internal
capacitances each time the impedance of the UCs needs to be reconfigured, and
the internal communication among the ASICs. Hence, by denoting the number of
reconfigurations of UC $i$ in a time window $T$ (this can be the frame duration)
by $N_{{\rm{rec}}_i}$ and the energy cost for such reconfigurations by
$E_{{\rm{rec}}_i}$, for the dynamic power consumption $P_{\rm{dyn}}$ of the RIS
it holds \cite[Eq. (4.7)]{D_5_3_VISORSURF}
\begin{equation}
	P_{\rm{dyn}}=\sum_{i \in
		\text{UCs}}\frac{N_{{\rm{rec}}_i}E_{{\rm{rec}}_i}}{T}.
\end{equation}

\begin{figure}
	\label{Frame_structure_time_UC_splitting}
	\centering
	{\includegraphics[width=2.6in,
		height=2in]{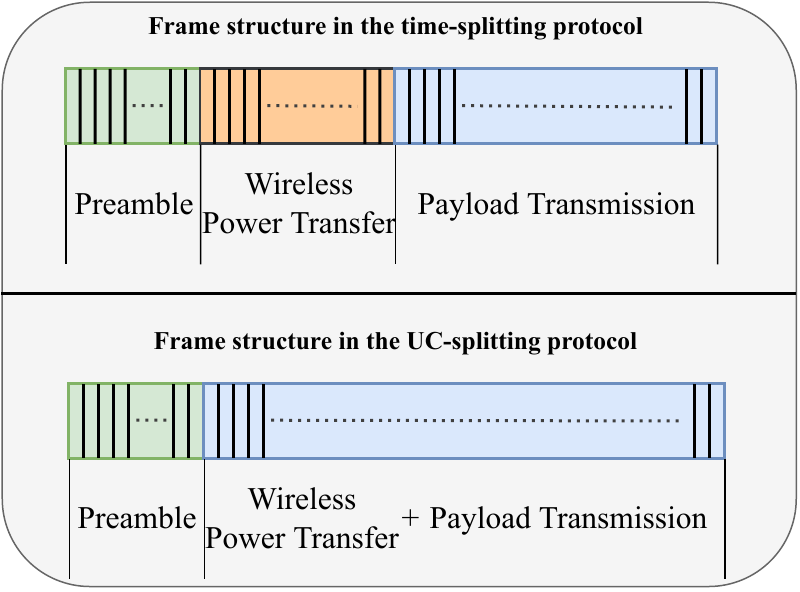}}
	\caption{Frame structure in the time- and UC-splitting harvesting protocols.}
	\label{Frame_structure_time_UC_splitting}	
\end{figure}

\begin{figure}
	\label{Time_splitting_intervals}
	\centering
	{\includegraphics[width=2.6in,
		height=3in]{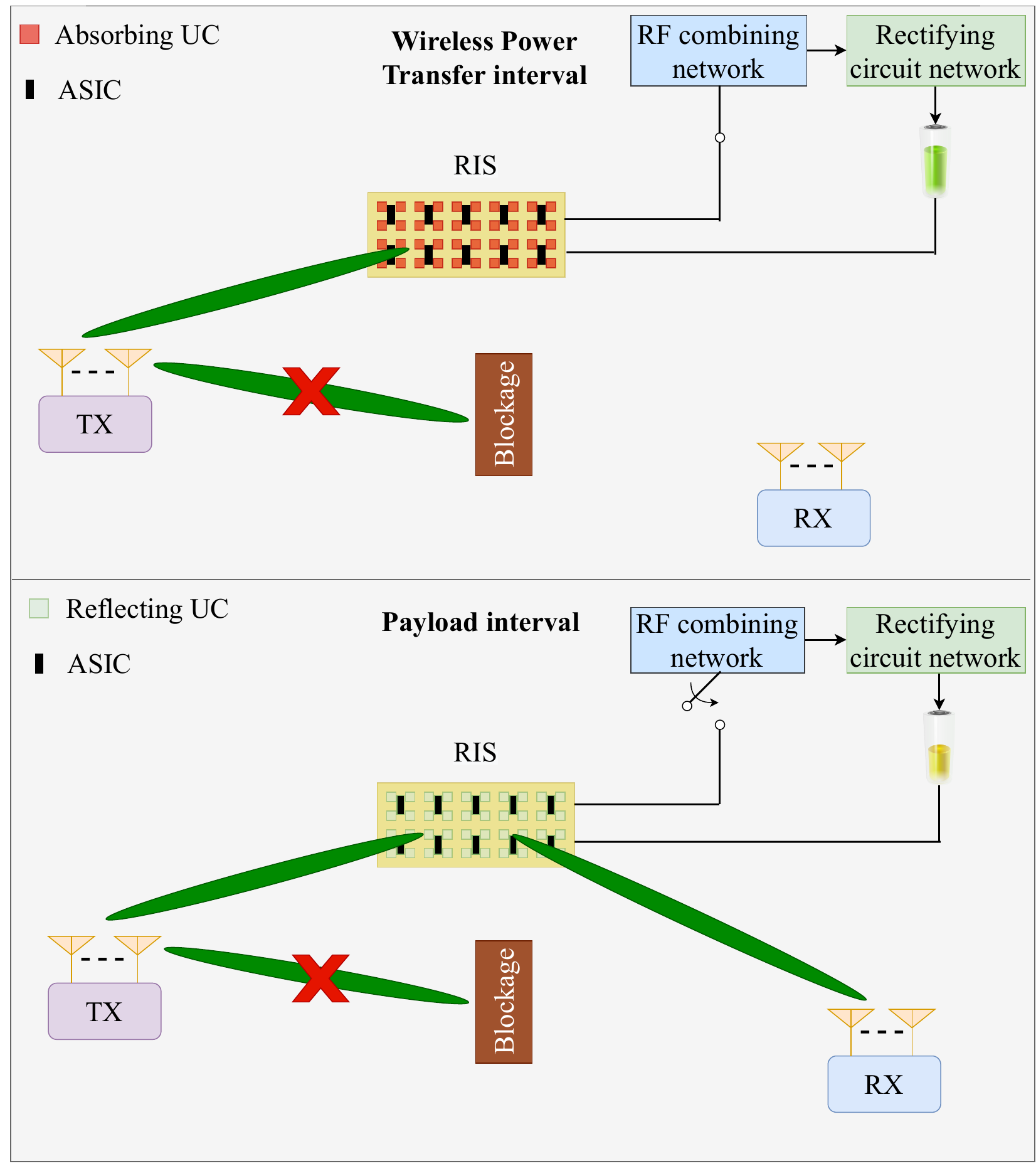}}
	\caption{Post-preamble time-splitting protocol functionality.}
	\label{Time_splitting_intervals}	
\end{figure}

On the other hand, the static power consumption is mainly attributed to the power consumption of the DACs, as \cite{Julius_ASIC_Low_Power} reveals.

\subsection*{Harvesting Protocols}
We now report two protocols for energy harvesting that are based on either a
time-splitting or a UC-splitting approach \cite{Time_vs_UC_splitting}.

\subsubsection*{Time-splitting protocol}
A typical frame structure is depicted in
Fig.~\ref{Frame_structure_time_UC_splitting}. Based on it, the preamble
interval, which is used for both synchronization and channel estimation of the
TX-RIS and RIS-RX links, is followed by an energy harvesting interval in which
all the UCs act as perfect absorbers. Finally, the payload transmission interval
follows where all the UCs act as perfect reflectors towards the RX. The
post-preamble functionality of the RIS is illustrated in
Fig.~\ref{Time_splitting_intervals}.

Let us now denote the number of UCs in the RIS by $M_s$. Regarding the number of
UC impedance adjustments that are needed during each frame, apart from $M_s$
adjustments needed for power absorption and another $M_s$ adjustments for the
payload transmission, based on the channel estimates, a number of UC adjustments
is needed for channel estimation during the preamble interval. The reason for
this becomes clear by considering that the RIS does not have active components
to perform channel estimation in order
to keep its design as simple and low-energy consuming as possible. Hence,
channel estimation involves only estimation at either the TX or RX. The simplest protocol for channel estimation relies on activating only one UC at a time
to act as perfect reflector while keeping the remaining ones off \cite{Zegrar_2021_EiD}. Hence, such a
channel estimation protocol requires in total $M_s$ UC impedance adjustments.
Based on the above, during the transmission of one frame in total $3M_s$ UC
reconfigurations are needed for channel estimation, wireless power absorption,
and payload transmission.

\subsubsection*{UC-splitting protocol}

\begin{figure}
	\label{Power_splitting_intervals}
	\centering
	{\includegraphics[width=2.6in,
		height=1.5in]{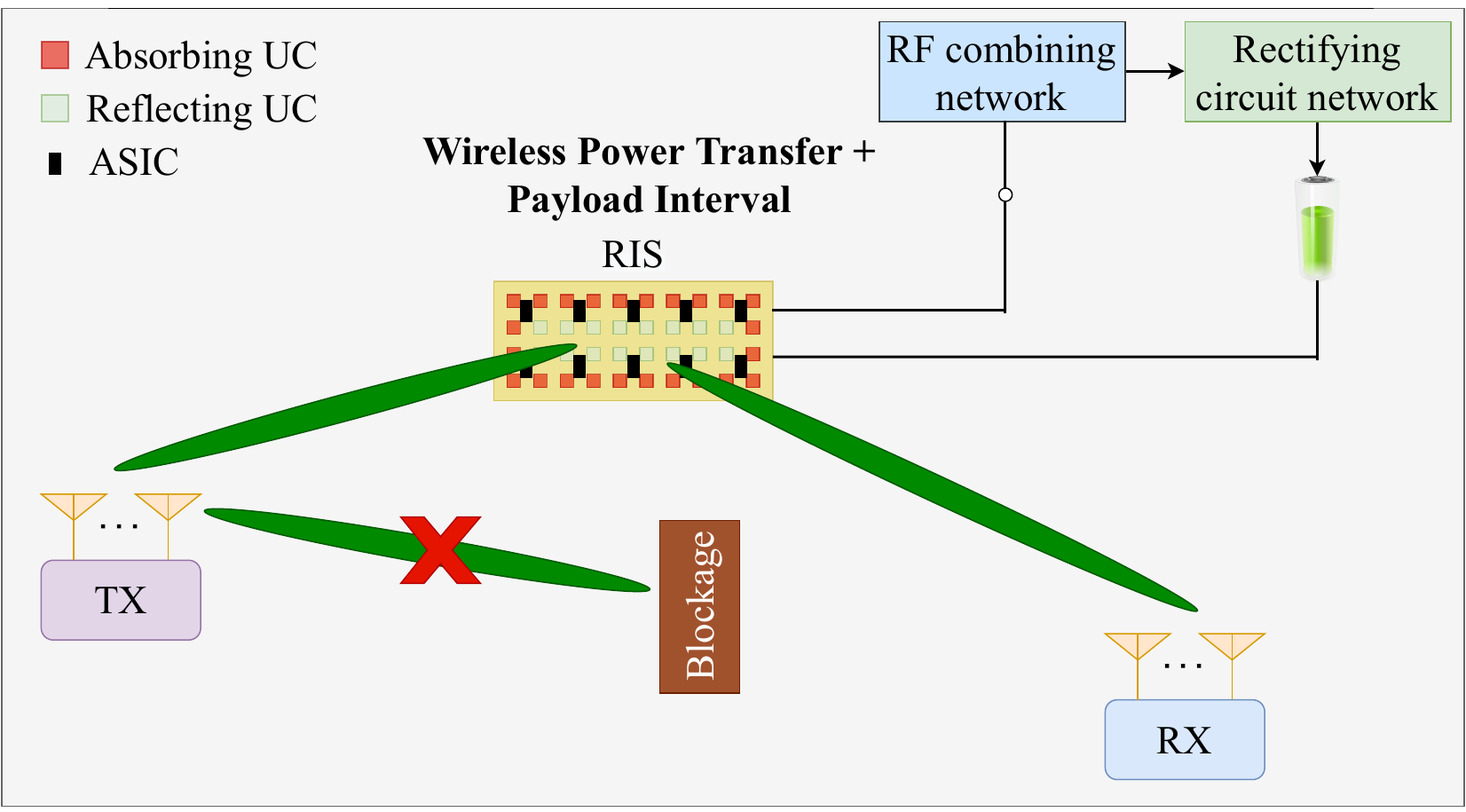}}
	\caption{Post-preamble power-splitting protocol functionality.}
	\label{Power_splitting_intervals}	
\end{figure}

The frame structure is depicted in
Fig.~\ref{Frame_structure_time_UC_splitting}. After the preamble transmission,
simultaneous wireless power transfer and information transmission is realized by
dedicating a subset of UCs for harvesting through perfect absorption and the
rest for information transmission by acting as perfect reflectors towards the
RX. Illustratively, the functionality of the RIS for the post-preamble frame
interval is depicted in Fig.~\ref{Power_splitting_intervals}.

Regarding the total number of UC reconfigurations needed in the UC-splitting
protocol during the transmission of one frame, $M_s$ reconfigurations are needed
for channel estimation and another $M_s$ reconfigurations for impedance
adjustment related to the simultaneous wireless power transfer and payload
transmission interval. Hence, $2M_s$ reconfigurations are needed in total, which
are smaller by $M_s$ reconfigurations compared with the time-splitting case.

Finally, we note that for the allocation of the time and UC resources in the
time- and UC-splitting harvesting protocols, respectively, average metrics can
be considered as the easiest implementation so that the allocation does not
depend on instantaneous channel estimates, but only on the channel statistics.

\section*{Performance Comparison of the Time and UC-Splitting protocols}

\begin{table}[t]
	\caption{Parameter values used in the simulations.\label{Parameter_values}} %
	\centering 
	\scalebox{0.8}{
		\begin{tabular}{| c | c | | c | c | } 
			\hline
			\textbf{Parameter} & \textbf{Value} & \textbf{Parameter} &
			\textbf{Value}\\[0.5ex]
			\hline
			\hline
			\text{Carrier frequency}& $28$ GHz& \thead{\text{UC number in the
					x-}\\\text{and y-axis of the RIS}} & $15$ \\[0.5ex] 
			\hline
			\text{Transmit power}& 1 W & \text{Bandwidth} & 200 MHz\\ [0.5ex]
			\hline
			\thead{\text{Transmit antenna }\\\text{gain}}& 37 dBi &	\thead{\text{Receive
					antenna }\\\text{gain}}& 24 dBi \\ [0.5ex]
			\hline
			\thead{\text{Transmitter-RIS}\\\text{distance}}& 18 m&
			\thead{\text{RIS-receiver}\\\text{distance}} &  38 m  \\ [0.5ex]
			\hline
			\thead{\text{Receiver noise figure}\\\text{gain}}& 10 dB &
			\thead{\text{Transmit/Receive}\\\text{antenna efficiency}} & 0.9 \\ [0.5ex]
			\hline
			\thead{\text{Time slot }\\\text{duration}}& 2 $\mu$s
			\cite{Julius_ASIC_Low_Power} & \thead{\text{Energy cost of}\\\text{an ASIC
					reconfiguration}} & 8 nJ \cite{Julius_ASIC_Low_Power} \\ [0.5ex]
			\hline
			\thead{\text{Transmitter-RIS}\\\text{channel}}& 
			\thead{\text{Free-space}\\\text{channel}}&\thead{\text{RIS-receiver}\\\text{channel}}&\thead{\text{Rician
					(K-factor}\\\text{equal to 10)}} \\[0.5ex]
			\hline
			\text{Preamble duration}& \thead{\text{$10^3$ time }\\\text{slots}}  &
			\text{Frame duration} & \thead{\text{$10^4$ time }\\\text{slots}} \\ [0.5ex]
			\hline
	\end{tabular}}
\end{table}


Let us now indicatively compare the performance of the time- and UC-splitting protocols in a typical communications-only scenario in which a mobile user is targeted via an RIS and the average rate maximization is the metric of interest. The simulation parameters are presented in Table~\ref{Parameter_values}. In addition, the energy-harvesting model and the harvesting circuit parameters of \cite{IEEE_TGCN_Energy_harvesting_RISs} are employed.

\begin{figure}
	\centering
	{\includegraphics[width=\columnwidth]{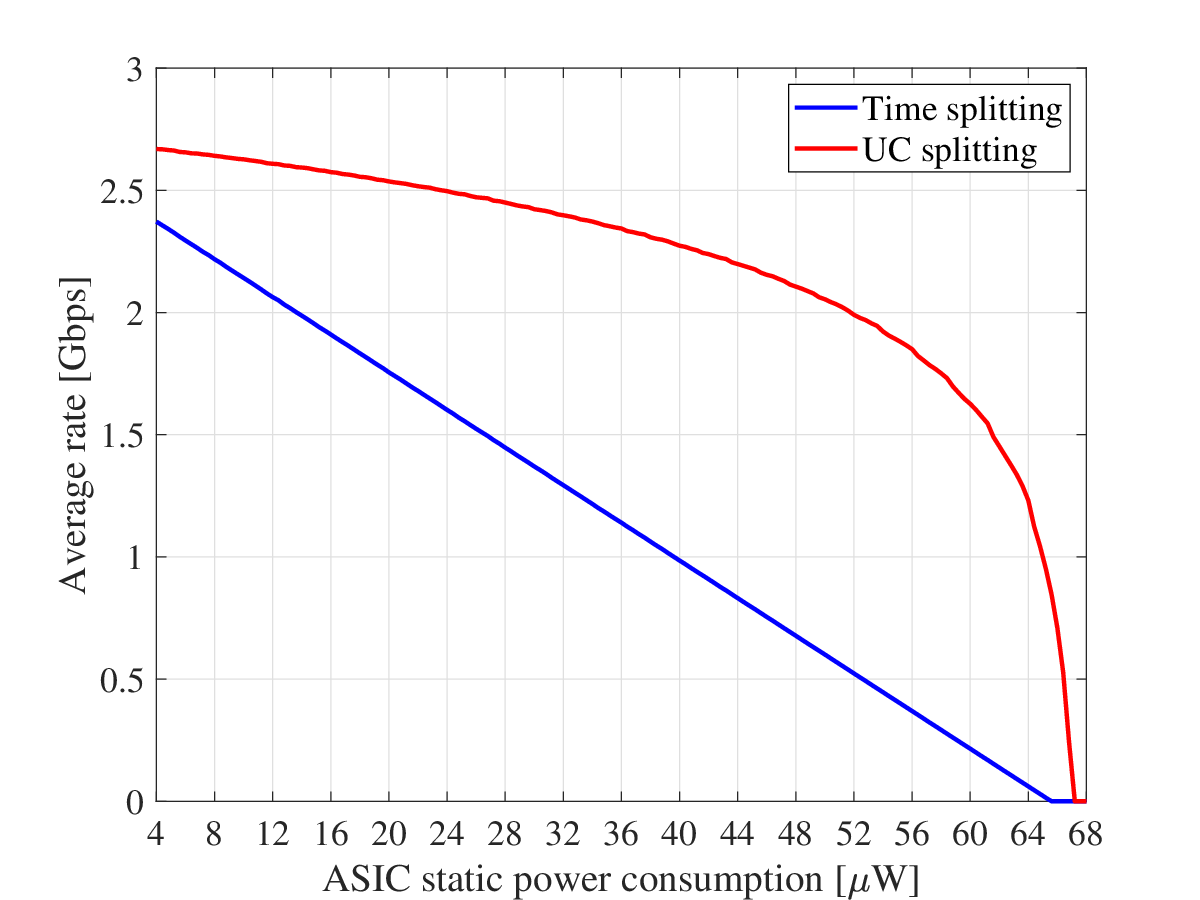}}
	\caption{Average rate vs. ASIC static power consumption.}
	\label{Performance_vs_RIS_power_consumption}	
\end{figure}

As far as the problem of the optimal resource allocation for the time- and UC-splitting protocols, we target the maximization of the average rate provided that the energy consumption requirements of the RIS are covered by the harvested energy. For the time-splitting protocol, such a problem takes the following form:

\begin{align}
	\begin{array}{l l}
		\underset{\thead{\text{Wireless power}\\\text{transfer duration} }}{\mathrm{maximize}} & \text{Average rate} \\
		\mathrm{subject\,\,to} & \thead{\text{DC harvested}\\ \text{power}}\ge \thead{\text{RIS power} \\ \text{consumption}}.
	\end{array}
\end{align}

On the other hand, in the case of the UC-splitting protocol the formulation of the optimal resource allocation problem is as follows:

\begin{align}
	\begin{array}{l l}
		\underset{\thead{\text{Number of UCs}\\\text{dedicated to} \\\text{energy harvesting}}}{\mathrm{maximize}} & \text{Average rate} \\ 
		\mathrm{subject\,\,to} & \thead{\text{DC harvested}\\ \text{power}}\ge \thead{\text{RIS power} \\ \text{consumption}}.
	\end{array}
\end{align}

Based on the solution of the presented problems, in Fig.~\ref{Performance_vs_RIS_power_consumption} we illustrate the average rate vs. the static ASIC power consumption that is achieved by the two protocols. The depicted ASIC static power consumption range is in the order of the one achieved in \cite{Julius_ASIC_Low_Power}. As we observe, in terms of average rate the UC-splitting protocol notably outperforms its time-splitting counterpart throughout the ASIC static power consumption range for which the solution of the two problems is feasible. This trend is justified by the fact that in the time-splitting case the factor corresponding to the reduction of time resources is a linear  multiplicative factor of Shannon's formula. On the other hand, for the UC-splitting protocol case such a term is included inside the logarithm function of Shannon's formula (in the signal-to-noise ratio (SNR) expression) \cite{IEEE_TGCN_Energy_harvesting_RISs}.

\begin{figure}
	\label{Dynamic_static_ASIC_power_ratio}
	\centering
	{\includegraphics[width=\columnwidth]{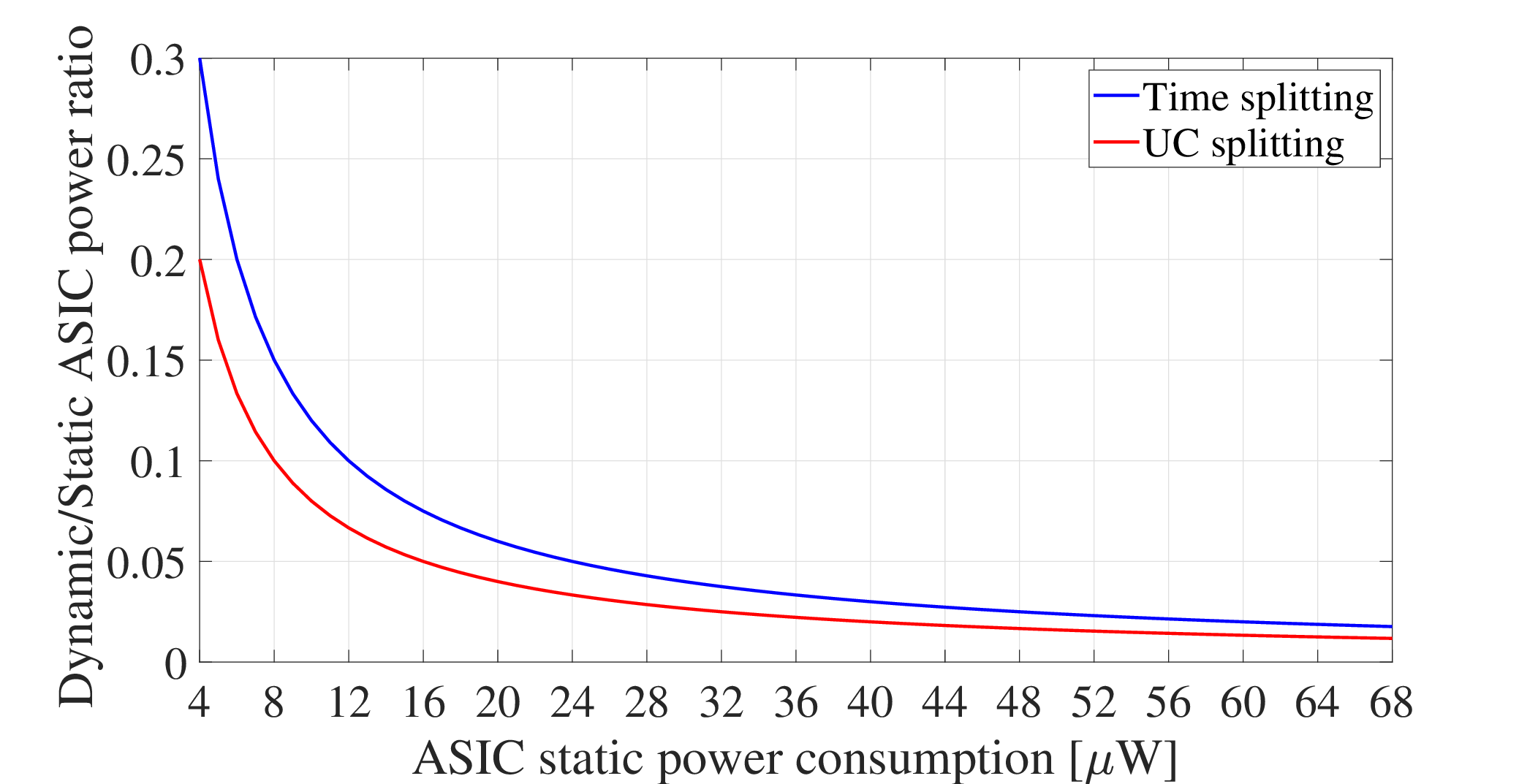}}
	\caption{Dynamic over static ASIC power consumption.}
	\label{Dynamic_static_ASIC_power_ratio}	
\end{figure}

Finally, it is interesting to examine the ratio of ASIC dynamic power consumption over the static one for the two examined protocols. This is depicted in Fig.~\ref{Dynamic_static_ASIC_power_ratio}. As we observe, as the ASIC static power consumption increases it largely dominates over the dynamic part. This is a clear indication that the realization of perpetual RISs dictates the design of ASICs that exhibit a very low static power consumption.
\section*{Challenges}

In this section, we present the main research challenges regarding the
realization of perpetual RISs and their deployment in future networks.

\subsection*{Low-Energy Consumption ASIC Design}

A key feature in the feasibility of perpetually operating RISs is the design of
ASICs that exhibit a very low static power consumption, as the simulation
results revealed. This is arquably the greatest obstacle to overcome. According
to the indicative simulation results, we saw that the ASICs of the RIS should
not consume more than just few $\mu$W of static power consumption for the
perpetual operation to be feasible. Instead, in the literature we observe that
typical ASICs used in integrated architecture designs exhibit a static power
consumption of few hundred of $\mu$W, which would render the perpetual RIS
operation infeasible \cite{Julius_ASIC_Low_Power}. More specifically, the most
power consuming component of the ASICs considered in
\cite{Julius_ASIC_Low_Power} is the DACs. In
addition, apart from the static power consumption per DAC, the number of DACs
and the number of UCs that each ASIC controls can be optimized so that the
perpetual operation is realized, based on the estimated amount of impinging
power.

\subsection*{Optimized Protocol Design for Energy Harvesting}

We have proposed two protocol architectures for RIS energy harvesting, namely the
time- and UC-splitting achitectures. As we saw in the previous section, the
latter architecture achieves a higher communication rate at the cost of a
reduced SNR, as revealed in \cite{Time_vs_UC_splitting},  since a portion of the UCs is dedicated to energy harvesting while
the rest simultaneously convey information. On the other hand, the
time-splitting architecture achieves the maximum SNR since all the UCs are
dedicated to the transmission of information. Besides this, having a relatively
high SNR at the receiver would be also important for the localization accuracy.
Hence, there should be a novel investigation of the most suitable energy
harvesting architecture for facilitating the demands of both communication and
localization. Most likely a stand-alone time- or UC-splitting architecture would
not be the way forward, but a dynamic switching between the time- and
UC-splitting architectures, depending on real-time demands would be needed in
real-world scenarios if it can be supported by the hardware.

\subsection*{Channel Modeling for Various High Frequency Bands}

Suitable high-frequency bands for all three purposes of RIS energy harvesting,
communication, and localization, is another innovative concept to investigate. In particular, it is known that due to electronics energy
harvesting becomes less efficient when going up the spectrum. However, very high
frequency bands, such as THz bands, offer the advantage of a stronger LoS
component due to the more directional transmissions and also finer resolution
for localization due to the larger bandwidths. In addition, the multipath
components that can also be harvested and importantly contribute to the absorbed
energy on the RIS, apart from the direct LoS component, can importantly add to
the required energy for supplying the energy needs of the RISs
\cite{IEEE_TGCN_Energy_harvesting_RISs}. Hence, accurate channel models for the
different high frequency bands are required. These aspects create very
interesting tradeoffs regarding the potential of different frequency bands for energy harvesting that need to be
investigated.

\subsection*{Network Planning}

The particular network planning will be based on
achieving the requirements on communications and localization with a certain
reliability, while at the same time the probability of not covering the RIS
energy demands is lower than a certain threshold. For such a network planning
reliable traffic models in a region are essential since these would determine
the statistical availability of the small cells for supplying the energy needs
of the RISs. For instance, apart from the energy supply that an RIS can receive
during the information transmission of its associated small cells, other,
possibly underutilized, small cells in that time instant could act as power
beacons for adding to the total harvested energy by the RISs.

\subsection*{Multi-Band Energy Harvesting}

The in-band energy-harvesting case examined in this work can be considered as a lower-bound scenario regarding the system performance considering that as the cost and size of electronics reduces eventually a perpetual RIS can host multi-band circuitry for energy harvesting. For instance, even in 6G and beyond networks that will mostly rely on mmWave bands for communication and localization, sub-6 GHz bands will still exist in multi-band small cells as a backup solution and also as a prime solution for control signals towards the mobile users. Hence, an RIS could incorporate both mmWave and sub-6 GHz circuitry to capture the ambient RF energy in the latter case from the small-cell transmissions. Additionally, another added energy-harvesting layer could relate to capturing solar energy in outdoor scenarios. Hence, the potential of multi-band energy harvesting should be investigated, taking also into account the cost and size of the resulting structure.

\subsection*{Communication- and Information-Theoretic Fundamental Limits}

The possibility of random energy arrivals in the case of multi-band energy harvesting, on top of the deterministic in-band harvesting that has been presented in this article, creates unique communication- and information-theoretic problems to be solved. Apart from the fact that in the presence of a ubiquitous RIS deployment the communication channel becomes programmable, with the existence of perpetual RISs the extent of its programmability depends on a random process that is related to the energy arrivals. From an information-theoretic point of view, a very interesting and challenging problem is the computation of the capacity of such a channel under finite-size batteries. In addition, channel coding theorems are of importance for such a novel system. Moreover, from a communications point of view, there is a need for practical adaptive modulation and coding schemes.

\subsection*{Real-Time Network Optimization}

Accurate analytical models for optimizing the resources in large-scale networks
that incorporate perpetual RISs would be intractable to obtain. This is due to
the complexity increase with respect to conventional networks that rely on
power-grid supplied RISs. In particular, taking into account the real-time
energy demands of the RISs substantially increases the optimal resource allocation
complexity.  Hence, data-driven approaches can be leveraged for the optimization
of the available network resources. However, obtaining the massive amount of
real-time data for training in centralized servers with the required latency and
network energy consumption seems a daunting task. For alleviating this,
distributed artificial intelligence methods can be leveraged, but this alone may
not be adequate. Consequently, to effectively tackle this issue offline data for
training through the use of less reliable analytical models that rely, for
instance, on stochastic-geometry approaches, can be examined \cite{Model_Data_Driven}. This way the
amount of real-time training can be notably reduced.

\section*{Conclusions}

The idea of perpetual RISs through RF in-band energy harvesting has been introduced in this article.
For its realization, it was firstly explained why the integrated architecture is potentially 
the only viable enabling architecture. Subsequently, we presented a typical EH
architecture together with the time- and UC-splitting protocols for in-band EH. An
indicative performance comparison between these two protocols followed, under an
optimal allocation of resources for maximizing the average rate,  which revealed
that the UC-splitting protocol largely outperforms its time-splitting
counterpart. Moreover, it was revealed that the static power consumption would most likely be the main
part of the total ASIC power consumption. Finally, from a hardware, link-level, and network perspective, several challenges, together with enablers to overcome them, have been identified towards the realization of large-scale communication networks with
perpetual RISs.

\bibliographystyle{IEEEtran}
\footnotesize{
	\bibliography{IEEEabrv, references, references_1, references_2}

\begin{thebibliography}{10}
\providecommand{\url}[1]{#1}
\csname url@samestyle\endcsname
\providecommand{\newblock}{\relax}
\providecommand{\bibinfo}[2]{#2}
\providecommand{\BIBentrySTDinterwordspacing}{\spaceskip=0pt\relax}
\providecommand{\BIBentryALTinterwordstretchfactor}{4}
\providecommand{\BIBentryALTinterwordspacing}{\spaceskip=\fontdimen2\font plus
\BIBentryALTinterwordstretchfactor\fontdimen3\font minus
  \fontdimen4\font\relax}
\providecommand{\BIBforeignlanguage}[2]{{%
\expandafter\ifx\csname l@#1\endcsname\relax
\typeout{** WARNING: IEEEtran.bst: No hyphenation pattern has been}%
\typeout{** loaded for the language `#1'. Using the pattern for}%
\typeout{** the default language instead.}%
\else
\language=\csname l@#1\endcsname
\fi
#2}}
\providecommand{\BIBdecl}{\relax}
\BIBdecl

\bibitem{Requirements_autonomous_vehicle_navigation}
T.~G.~R. Reid \emph{et~al.}, ``{Localization Requirements for Autonomous
  Vehicles},'' \emph{SAE nternational Journal of Connected and Automated
  Vehicles}, June 2019.

\bibitem{Overview_Integrated_Localization_Paper}
Z.~{Xiao} and Y.~{Zeng}, ``{An Overview on Integrated Localization and
  Communication Towards 6G},'' \emph{arXiv:2006.01535}, Jun. 2020.

\bibitem{Indoor_Navigation_Visual_Impairments}
D.~Ahmetovic \emph{et~al.}, ``Achieving practical and accurate indoor
  navigation for people with visual impairments,'' in \emph{W4A '17:
  Proceedings of the 14th International Web for All Conference}, April 2017.

\bibitem{Khawaja_mmWave_Passive_Reflectors}
W.~{Khawaja}, O.~{Ozdemir}, Y.~{Yapici}, F.~{Erden}, and I.~{Guvenc},
  ``Coverage {E}nhancement for {NLOS} mmwave {L}inks {U}sing {P}assive
  {R}eflectors,'' \emph{IEEE Open J. Commun. Soc.}, vol.~1, no.~1, pp.
  263--281, Jan. 2020.

\bibitem{Cunhua_RIS_6G}
C.~Pan, H.~Ren, K.~Wang, J.~F. Kolb, M.~Elkashlan, M.~Chen, M.~Di~Renzo,
  Y.~Hao, J.~Wang, A.~L. Swindlehurst, X.~You, and L.~Hanzo, ``{Reconfigurable
  Intelligent Surfaces for 6G Systems: Principles, Applications, and Research
  Directions},'' \emph{IEEE Communications Magazine}, vol.~59, no.~6, pp.
  14--20, 2021.

\bibitem{Jiguang_He_RIS_Simultaneous_Communications_Localization}
J.~He, F.~Jiang, K.~Keykhosravi, J.~Kokkoniemi, H.~Wymeersch, and M.~Juntti,
  ``{Beyond 5G RIS mmWave Systems: Where Communication and Localization
  Meet},'' \emph{IEEE Access}, vol.~10, pp. 68\,075--68\,084, 2022.

\bibitem{Flexibility_cost_energy_efficiency_smart_home}
\BIBentryALTinterwordspacing
``{Flexibility Cost and Energy Efficiency in Smart Home},'' Tech. Rep., 2015.
  [Online]. Available:
  \url{https://switches-sensors.zf.com/us/flexibility-cost-energy-efficiency-in-smart-home/}
\BIBentrySTDinterwordspacing

\bibitem{Abadal_2020_programmable_metamaterials}
S.~{Abadal}, T.~{Cui}, T.~{Low}, and J.~{Georgiou}, ``Programmable
  metamaterials for software-defined electromagnetic control: Circuits,
  systems, and architectures,'' \emph{IEEE Journal on Emerging and Selected
  Topics in Circuits and Systems}, vol.~10, no.~1, pp. 6--19, 2020.

\bibitem{Julius_ASIC_Low_Power}
L.~Petrou and J.~Georgiou, ``{An ASIC Architecture With Inter-Chip Networking
  for Individual Control of Adaptive-Metamaterial Cells},'' \emph{IEEE Access},
  vol.~10, pp. 80\,234--80\,248, 2022.

\bibitem{Energy_harvesting_metasurface_comprehensive_review}
A.~A.~G. Amer, S.~Z. Sapuan, N.~Nasimuddin, A.~Alphones, and N.~B. Zinal, ``{A
  Comprehensive Review of Metasurface Structures Suitable for RF Energy
  Harvesting},'' \emph{IEEE Access}, vol.~8, pp. 76\,433--76\,452, 2020.

\bibitem{D_5_3_VISORSURF}
\BIBentryALTinterwordspacing
``{Report on the comparison between ideal HyperSurface {(HSF)}s and the
  manufactured prototypes},'' VISORSURF project, Tech. Rep., Dec. 2020.
  [Online]. Available:
  \url{https://ec.europa.eu/research/participants/documents/downloadPublic?documentIds=080166e5d7993c7d&appId=PPGMS}
\BIBentrySTDinterwordspacing

\bibitem{Time_vs_UC_splitting}
K.~Ntontin, E.~Björnson, A.~A. Boulogeorgos, Z.~Abdullah, A.~Mesodiakaki,
  S.~Abadal, and S.~Chatzinotas, ``Time-and unit-cell splitting comparison for
  the autonomous operation of reconfigurable intelligent surfaces,'' \emph{IEEE
  Transactions on Green Communications and Networking}, to appear.

\bibitem{Zegrar_2021_EiD}
S.~Eddine~Zegrar, L.~Afeef, and H.~Arslan, ``{Reconfigurable intelligent
  surface (RIS): Eigenvalue Decomposition-Based Separate Channel Estimation},''
  in \emph{2021 IEEE 32nd Annual International Symposium on Personal, Indoor
  and Mobile Radio Communications (PIMRC)}, 2021.

\bibitem{IEEE_TGCN_Energy_harvesting_RISs}
K.~Ntontin, A.-A.~A. Boulogeorgos, E.~Bjornson, W.~A. Martins, S.~Kisseleff,
  S.~Abadal, E.~Alarcon, A.~Papazafeiropoulos, F.~Lazarakis, and
  S.~Chatzinotas, ``{Wireless Energy Harvesting For Autonomous Reconfigurable
  Intelligent Surfaces},'' \emph{IEEE Transactions on Green Communications and
  Networking}, vol.~7, no.~1, pp. 114--129, March 2023.

\bibitem{Model_Data_Driven}
A.~Zappone, M.~Di~Renzo, and M.~Debbah, ``Wireless networks design in the era
  of deep learning: Model-based, ai-based, or both?'' \emph{IEEE Transactions
  on Communications}, vol.~67, no.~10, pp. 7331--7376, 2019.

\end{thebibliography}
}

\end{document}